\documentclass[pdftex,twocolumn,epjc3]{svjour3}

\RequirePackage[utf8]{inputenc}
\RequirePackage{color}
\RequirePackage{graphicx}
\RequirePackage{mathptmx}      % use Times fonts if available on your TeX system
\RequirePackage{flushend}
\RequirePackage[numbers,sort&compress]{natbib}
\RequirePackage[colorlinks,citecolor=blue,urlcolor=blue,linkcolor=blue]{hyperref}
\RequirePackage{amsmath}
\RequirePackage{amssymb}
\RequirePackage{enumitem}
\RequirePackage{lineno}

\DeclareMathAlphabet{\mathcal}{OMS}{cmsy}{m}{n}

\def\d{{\rm d}}
\def\un#1{\,{\rm #1}}

\def\unt#1{[{\rm #1}]}

\def\text#1{\hbox{#1}}

\setbox123\hbox{\small$0$}
\def\S{\hbox to\wd123{\hss}}
\setbox124\hbox{\small$_{0}$}
\def\s{\hbox to\wd124{\hss}}

\journalname{Eur. Phys. J. C}

\begin{document}

\title{Characterisation of the dip-bump structure observed in proton-proton elastic scattering at $\sqrt s$ = 8\,TeV}
\titlerunning{Characterisation of the dip-bump structure observed in proton-proton elastic scattering at $\sqrt s$ = 8\,TeV}

%----------------------------------------------------------------------------------------------------
% Add an author
% USAGE: \AddAuthor{name}{primary affiliation}{secondary affiliation}{external affiliation}

\newif\ifFirstAuthor
\FirstAuthortrue

\def\AddAuthor#1#2#3#4{%
	\def\PriAf{#2}%
	\def\SecAf{#3}%
	\def\ExtAf{#4}%
	\def\empty{}%
	\ifFirstAuthor
		\FirstAuthorfalse
	\else
		\and
	\fi
	\ifx\PriAf\empty
		% no primary affiliation, only external
		#1\thanksref{#4}%
	\else
		\ifx\SecAf\empty
			\ifx\ExtAf\empty
				% only primary affiliation
				#1\thanksref{#2}%
			\else
				% primary and external affiliation
				#1\thanksref{#2,#4}%
			\fi
		\else
			\ifx\ExtAf\empty
				% primary and secondary affiliation
				#1\thanksref{#2,#3}%
			\else
				% primary, secondary and external affiliation
				#1\thanksref{#2,#3,#4}%
			\fi
		\fi
	\fi
}

%----------------------------------------------------------------------------------------------------
% Add a TOTEM institute
% USAGE: \AddInstitute{reference as in AddAuthor}{address}

%	First Address, Street, City, Country\label{addr1} \and

\newif\ifFirstInstitute
\FirstInstitutetrue

\def\AddInstitute#1#2{%
	\ifFirstInstitute
		\FirstInstitutefalse
	\else
		\and
	\fi
	#2\label{#1}%
}

%----------------------------------------------------------------------------------------------------
% Add an institute not in the TOTEM collaboration
% USAGE: \AddExternalInstitute{reference as in AddAuthor}{address}

\def\AddExternalInstitute#1#2{%
	\thankstext{#1}{#2}
}

%----------------------------------------------------------------------------------------------------
%----------------------------------------------------------------------------------------------------
% include the actual data

\def\DeclareAuthors{%
	\AddAuthor{G.~Antchev}{}{}{1}%
	\AddAuthor{P.~Aspell}{9}{}{}%
	\AddAuthor{I.~Atanassov}{}{}{1}%
	\AddAuthor{V.~Avati}{7}{9}{}%
	\AddAuthor{J.~Baechler}{9}{}{}%
	\AddAuthor{C.~Baldenegro Barrera}{11}{}{}%
	\AddAuthor{V.~Berardi}{4a}{4b}{}%
	\AddAuthor{M.~Berretti}{2a}{}{}%
	\AddAuthor{V.~Borchsh}{8}{}{}%
	\AddAuthor{E.~Bossini}{6aa}{9}{}%
	\AddAuthor{U.~Bottigli}{6b}{}{}%
	\AddAuthor{M.~Bozzo}{5a}{5b}{}%
	\AddAuthor{H.~Burkhardt}{9}{}{}%
	\AddAuthor{F.~S.~Cafagna}{4a}{}{}%
	\AddAuthor{M.~G.~Catanesi}{4a}{}{}%
	\AddAuthor{M.~Csan\'{a}d}{3a}{}{2}%
	\AddAuthor{T.~Cs\"{o}rg\H{o}}{3a}{3b}{}%
	\AddAuthor{M.~Deile}{9}{}{}%
	\AddAuthor{F.~De~Leonardis}{4c}{4a}{}%
	\AddAuthor{M.~Doubek}{1c}{}{}%
	\AddAuthor{D.~Druzhkin}{8}{9}{}%
	\AddAuthor{K.~Eggert}{10}{}{}%
	\AddAuthor{V.~Eremin}{}{}{4}%
	\AddAuthor{F.~Ferro}{5a}{}{}%
	\AddAuthor{A.~Fiergolski}{9}{}{}%	
	\AddAuthor{L.~Forthomme}{2a}{2b}{}%
	\AddAuthor{F.~Garcia}{2a}{}{}%
	\AddAuthor{V.~Georgiev}{1a}{}{}%
	\AddAuthor{S.~Giani}{9}{}{}%
	\AddAuthor{L.~Grzanka}{7}{}{}%
	\AddAuthor{J.~Hammerbauer}{1a}{}{}%
	\AddAuthor{T.~Isidori}{11}{}{}%
	\AddAuthor{V.~Ivanchenko}{8}{}{}%
	\AddAuthor{M.~Janda}{1c}{}{}%
	\AddAuthor{A.~Karev}{9}{}{}%
	\AddAuthor{J.~Ka\v{s}par}{1b}{9}{}%
	\AddAuthor{B.~Kaynak}{}{}{5}%
	\AddAuthor{J.~Kopal}{9}{}{}%
	\AddAuthor{V.~Kundr\'{a}t}{1b}{}{}%
	\AddAuthor{S.~Lami}{6a}{}{}%
	\AddAuthor{G.~Latino}{6b}{}{}%
	\AddAuthor{R.~Linhart}{1a}{}{}%
	\AddAuthor{C.~Lindsey}{11}{}{}%
	\AddAuthor{M.~V.~Lokaj\'{\i}\v{c}ek}{1b}{}{dead}%
	\AddAuthor{L.~Losurdo}{6b}{}{}%
	\AddAuthor{F.~Lucas~Rodr\'{i}guez}{9}{}{}%
	\AddAuthor{M.~Macr\'{\i}}{5a}{}{}%
	\AddAuthor{M.~Malawski}{7}{}{}%
	\AddAuthor{N.~Minafra}{11}{}{}%
	\AddAuthor{S.~Minutoli}{5a}{}{}%
	\AddAuthor{T.~Naaranoja}{2a}{2b}{}%
	\AddAuthor{F.~Nemes}{9}{3a}{}%
	\AddAuthor{H.~Niewiadomski}{10}{}{}%
	\AddAuthor{T.~Nov\'{a}k}{3b}{}{}%
	\AddAuthor{E.~Oliveri}{9}{}{}%
	\AddAuthor{F.~Oljemark}{2a}{2b}{}%
	\AddAuthor{M.~Oriunno}{}{}{6}%
	\AddAuthor{K.~\"{O}sterberg}{2a}{2b}{}%
	\AddAuthor{P.~Palazzi}{9}{}{}%
	\AddAuthor{V.~Passaro}{4c}{4a}{}%
	\AddAuthor{Z.~Peroutka}{1a}{}{}%
	\AddAuthor{J.~Proch\'{a}zka}{1b}{}{}%
	\AddAuthor{M.~Quinto}{4a}{4b}{}%
	\AddAuthor{E.~Radermacher}{9}{}{}%
	\AddAuthor{E.~Radicioni}{4a}{}{}%
	\AddAuthor{F.~Ravotti}{9}{}{}%
	\AddAuthor{E.~Robutti}{5a}{}{}%
	\AddAuthor{C.~Royon}{11}{}{}%
	\AddAuthor{G.~Ruggiero}{9}{}{}%
	\AddAuthor{H.~Saarikko}{2a}{2b}{}%
	\AddAuthor{V.D.~Samoylenko}{}{}{3}%
	\AddAuthor{A.~Scribano}{6a}{}{}%
	\AddAuthor{J.~Siroky}{1a}{}{}%
	\AddAuthor{J.~Smajek}{9}{}{}%
	\AddAuthor{W.~Snoeys}{9}{}{}%
	\AddAuthor{R.~Stefanovitch}{9}{}{}%
	\AddAuthor{J.~Sziklai}{3a}{}{}%
	\AddAuthor{C.~Taylor}{10}{}{}%
	\AddAuthor{E.~Tcherniaev}{8}{}{}%
	\AddAuthor{N.~Turini}{6b}{}{}%
	\AddAuthor{O.~Urban}{1a}{}{}%
	\AddAuthor{V.~Vacek}{1c}{}{}%
	\AddAuthor{O.~Vavroch}{1a}{}{}%
	\AddAuthor{J.~Welti}{2a}{2b}{}%
	\AddAuthor{J.~Williams}{11}{}{}%
	\AddAuthor{J.~Zich}{1a}{}{}%
	\AddAuthor{K.~Zielinski}{7}{}{}%
}

%----------------------------------------------------------------------------------------------------

\def\DeclareInstitutes{%
	\AddInstitute{1a}{University of West Bohemia, Pilsen, Czech Republic.}
	\AddInstitute{1b}{Institute of Physics of the Academy of Sciences of the Czech Republic, Prague, Czech Republic.}
	\AddInstitute{1c}{Czech Technical University, Prague, Czech Republic.}
	\AddInstitute{2a}{Helsinki Institute of Physics, University of Helsinki, Helsinki, Finland.}
	\AddInstitute{2b}{Department of Physics, University of Helsinki, Helsinki, Finland.}
	\AddInstitute{3a}{Wigner Research Centre for Physics, RMI, Budapest, Hungary.}
	\AddInstitute{3b}{MATE Institute of Technology KRC, Gy\"ongy\"os, Hungary.}
	\AddInstitute{4a}{INFN Sezione di Bari, Bari, Italy.}
	\AddInstitute{4b}{Dipartimento Interateneo di Fisica di Bari, Bari, Italy.}
	\AddInstitute{4c}{Dipartimento di Ingegneria Elettrica e dell'Informazione - Politecnico di Bari, Bari, Italy.}
	\AddInstitute{5a}{INFN Sezione di Genova, Genova, Italy.}
	\AddInstitute{5b}{Universit\`{a} degli Studi di Genova, Italy.}
	\AddInstitute{6a}{INFN Sezione di Pisa, Pisa, Italy.}
	\AddInstitute{6aa}{Universit\`{a} degli Studi di Pisa, Pisa, Italy.}
	\AddInstitute{6b}{Universit\`{a} degli Studi di Siena and Gruppo Collegato INFN di Siena, Siena, Italy.}
	\AddInstitute{7}{AGH University of Science and Technology, Krakow, Poland.}
	\AddInstitute{8}{Tomsk State University, Tomsk, Russia.}
	\AddInstitute{9}{CERN, Geneva, Switzerland.}
	\AddInstitute{10}{Case Western Reserve University, Dept.~of Physics, Cleveland, OH, USA.}
	\AddInstitute{11}{The University of Kansas, Lawrence, USA.}
}

%----------------------------------------------------------------------------------------------------
	
\def\DeclareExternalInstitutes{%
	\AddExternalInstitute{1}{INRNE-BAS, Institute for Nuclear Research and Nuclear Energy, Bulgarian Academy of Sciences, Sofia, Bulgaria.}
	\AddExternalInstitute{2}{Department of Atomic Physics, ELTE University, Budapest, Hungary.}
	\AddExternalInstitute{3}{NRC `Kurchatov Institute'-IHEP, Protvino, Russia.}
	\AddExternalInstitute{4}{Ioffe Physical - Technical Institute of Russian Academy of Sciences, St.~Petersburg, Russian Federation.}
	\AddExternalInstitute{5}{Istanbul University, Istanbul, Turkey.}
	\AddExternalInstitute{6}{SLAC National Accelerator Laboratory, Stanford CA, USA.}
	\AddExternalInstitute{dead}{Deceased.}
}

\author{%
	The TOTEM Collaboration\\
	\DeclareAuthors
}

\DeclareExternalInstitutes

\institute{%
	\DeclareInstitutes
}

\authorrunning{The TOTEM Collaboration}

\date{Manuscript date: \today}

\maketitle

\begin{abstract}
The TOTEM collaboration at the CERN LHC has measured the differential cross-section of elastic proton-proton scattering at $\sqrt{s} = 8\un{TeV}$ in the squared four-momentum transfer range $0.2\un{GeV^{2}} < |t| < 1.9\un{GeV^{2}}$. This interval includes the structure with a diffractive minimum (``dip'') and a secondary maximum (``bump'') that has also been observed at all other LHC energies, where measurements were made. A detailed characterisation of this structure for $\sqrt{s} = 8\un{TeV}$ yields the positions,
$|t|_{\rm dip} = (0.521 \pm 0.007)\un{GeV^2}$
and
$|t|_{\rm bump} = (0.695 \pm 0.026)\un{GeV^2}$, as well as the cross-section values,
$\left.{\d\sigma/\d t}\right|_{\rm dip} = (15.1 \pm 2.5)\un{{\mu b/GeV^2}}$
and
$\left.{\d\sigma/\d t}\right|_{\rm bump} = (29.7 \pm 1.8)\un{{\mu b/GeV^2}}$, for the dip and the bump, respectively. 
\end{abstract}

% TODO: remove at the end
%\linenumbers

%----------------------------------------------------------------------------------------------------
\section{Introduction}
\label{s:introduction}

The TOTEM experiment at Interaction Point 5 (IP5) of the CERN Large Hadron Collider (LHC) has measured the differential cross-section $\d\sigma/\d t$ of elastic proton-proton scattering at a centre-of-mass energy $\sqrt{s} = 8\un{TeV}$ in the region of the structure with a diffractive minimum and a secondary maximum (a.k.a.~dip-bump) by extending an earlier analysis~\cite{totem-8tev-90m} of the same dataset to $|t|$-values up to $1.9\un{GeV^{2}}$, where $t$ is the squared four-momentum transfer. The new analysis reported in the present article completes the series of elastic cross-section measurements at all LHC energies reached in Runs 1 and 2: $\sqrt{s} = 2.76\un{TeV}$~\cite{totem-2.76tev-diff}, $7\un{TeV}$~\cite{totem-7tev-3.5m}, $8\un{TeV}$, $13\un{TeV}$~\cite{totem-13tev-diff}. This measurement series is of particular interest for the comparison between proton-proton (pp) and proton-antiproton (p$\bar{\rm p}$) scattering at the TeV energy scale. While in all pp datasets a very distinct dip-bump structure is observed, the only available p$\bar{\rm p}$ scattering measurement in the TeV energy range, performed at $\sqrt{s} = 1.96\un{TeV}$ by the Tevatron D0 experiment~\cite{d0paper}, exhibits only a shoulder. Having pp measurements at several different energies gives access to a quantitative characterisation of the dip-bump structure and its energy dependence. A detailed comparison of the elastic $\d\sigma/\d t$ between $\rm pp$ and $\rm p\bar p$ can be found in Ref.~\cite{d0-totem-prl}. The observed difference between the elastic $\d\sigma/\d t$ of $\rm pp$ and $\rm p\bar p$ scattering at the TeV scale points to the existence of a C-odd t-channel exchange, the Odderon, in addition to the dominant C-even exchange, the Pomeron, in elastic scattering.

The measurement reported here was carried out with the Roman Pot (RP) system, the TOTEM subdetector for leading protons~\cite{totem-jinst}. A Roman Pot is a beam-pipe insertion designed to move a detector -- a stack of 10 silicon sensor planes in the case of TOTEM -- towards the beam when the accelerator has reached stable beam conditions. Thus the tracking detectors can approach the beam centre to distances of the order of a millimetre and detect protons scattered at angles in the microradian range. In LHC Run 1, when the data for this article were collected, the RP system consisted of four units of Roman Pots installed at $\pm$214\,m and $\pm$220\,m from IP5 on the outgoing beamlines, i.e. in the LHC sectors 45 and 56. Each unit consists of three RPs: a vertical pair approaching the beam from the top and bottom, and an individual horizontal RP. 

The data were collected in July 2012 in the dedicated LHC fill \#2836 with a special beam optics where the betatron function in IP5 had the value $\beta^* = 90\un{m}$~\cite{90m-optics}. This configuration provided a small beam divergence and thus a good resolution in the scattering angle $\theta^{*}$ and hence in $t\approx - p^{2} \theta^{*\, 2}$, where $p$ is the beam momentum\footnote{The squared four-momentum transfer $t$ is always negative. Throughout this paper the modulus $|t|$ is used.}. It also had a large vertical effective length\footnote{Effective length: the optical function translating the scattering-angle at the IP into a displacement of the scattered proton from the beam centre at the RP.} yielding a good acceptance at low $|t|$ with the vertical RPs, inserted at a distance of $9.5$ times the transverse size of the beam, $\sigma_{\rm beam}$. More details of the optics are reported in Ref.~\cite{totem-8tev-90m}.

Since elastic-scattering events consist of two collinear protons emitted in opposite directions, the signal events can have
two topologies, called ``diagonals'': 45 bottom -- 56 top and 45 top -- 56 bottom. The main trigger required a coincidence between the RPs in both arms. During the about $11\un{h}$ long data-taking, a luminosity of about $735\un{\mu b^{-1}}$ was accumulated.

%----------------------------------------------------------------------------------------------------
\section{Differential cross-section}
\label{s:dsdt}

The analysis procedure is almost identical to the one published in Ref.~\cite{totem-8tev-90m}. Here only a brief overview is given, for details the reader is referred to the original publication.

For a given $t$ bin, the differential cross-section is evaluated by selecting and counting elastic events:
\begin{equation}
	{\d\sigma\over \d t}\left(\hbox{bin}\right) =
		{\cal N}\: {\cal U}(t)\: {\cal B} \: 
		\frac{1}{\Delta t}
                \sum\limits_{t\, \in\, \text{bin}} {\cal A}(t, t_y)\: {\cal E}(t_y) \: ,
\end{equation}
where $\Delta t$ is the width of the bin, ${\cal N}$ is a normalisation factor, 
and the other symbols stand for correction factors:
${\cal U}$ for unfolding of resolution effects, ${\cal B}$ for background subtraction, ${\cal A}$ for acceptance correction and ${\cal E}$ for detection and reconstruction efficiency. $t_y \equiv - p^{2} \theta_{y}^{*\, 2}$ represents the component of the four-momentum transfer squared related to the vertical scattering angle, relevant for some of the corrections.

The candidate events are tagged with cuts that enforce the elastic-event kinematics: two collinear protons (one in each arm of the experiment) emerging from the same vertex. In addition, the optics-imposed correlation between the vertical track position and angle at the RPs is required. All the cuts are applied at the $4\un{\sigma}$ level where Monte Carlo studies indicate a tolerable loss of about $0.1\un{\%}$ of the events.

The background, i.e.~non-elastic events passing the tagging cuts, is determined  by analysing the distributions of several selection discriminators (e.g.~the difference between the reconstructed scattering angles from the two arms, $\theta_{x,y}^{*\,56} - \theta_{x,y}^{*\,45}$, see Table~2 in Ref.~\cite{totem-8tev-90m}) for two complementary data sets: (a) the events with diagonal topology, containing both elastic signal and non-elastic background, and (b) the events with anti-diagonal topology (i.e. 45 top -- 56 top, or 45 bottom -- 56 bottom), which cannot contain any elastic signal. For the diagonal events, the tails of the discriminator distributions, containing only background, are interpolated into the signal region to estimate the contamination of that region. The shape used for this interpolation is taken from the anti-diagonal events. In the tail region, the discriminator distributions of the anti-diagonal and the diagonal events have been confirmed to agree well. Hence it is expected that also in the signal regions of the discriminators the distributions of the anti-diagonal events are similar to the background part of the diagonal events. This procedure yields a background estimate of $1 - {\cal B} < 10^{-4}$.

%The first is based on distributions of the selection discriminators -- the tails contain only background, which can be interpolated into the signal region. The second method determines the shape for this interpolation using the data from anti-diagonal configurations (i.e. 45 top -- 56 top, or 45 bottom -- 56 bottom), which cannot contain any elastic signal but where the background is expected to be similar. This expectation is confirmed by the good agreement between the diagonal and anti-diagonal configurations in the distribution tails. This procedure yields a background estimate of $1 - {\cal B} < 10^{-4}$.

The acceptance correction, ${\cal A}$, receives two contributions. The ``geometrical'' correction reflects the fraction of events with a given value of $|t|$ that fall within the geometrical acceptance of the sensors. The second contribution corrects for fluctuations around the sensor edges mainly due to the beam divergence.

The normalisation, ${\cal N}$, is determined by requiring the same cross-section integral between $|t| = 0.027$ and $0.083\un{GeV^2}$ as for dataset 1 from Ref.~\cite{totem-8tev-tot1}, where the luminosity-independent calibration was applied.

Since the normalisation is determined from another dataset, in the present analysis it is sufficient to consider only inefficiency effects, ${\cal E}$, that may modify the $t$-distribution shape. These are caused by the inability of the system to reconstruct an elastic proton track. Two cases are distinguished. In the first case, a single RP does not show one unique proton track (it may have either zero or several tracks, which cannot be resolved in a strip detector system). Such inefficiencies are evaluated by removing the RP from the tagging cuts, repeating the selection and calculating the fraction of events recovered. In the second case, multiple RPs in the same arm do not show the proton track, which typically results from showers, initiated in the upstream RP and affecting also the downstream one. The related inefficiency is studied by examining the rate of events with high track multiplicity.

The scattering-angle resolution is studied by comparing the protons in the two arms of the RP system. For elastic events the angles should be identical, but fluctuations arise due to the beam divergence and partly due to the finite RP sensor resolution. The scattering-angle resolution was found to deteriorate slightly with time, at a rate compatible with the beam emittance growth.

\begin{figure}
%\hbox{}\vskip-7mm
\begin{center}
\includegraphics{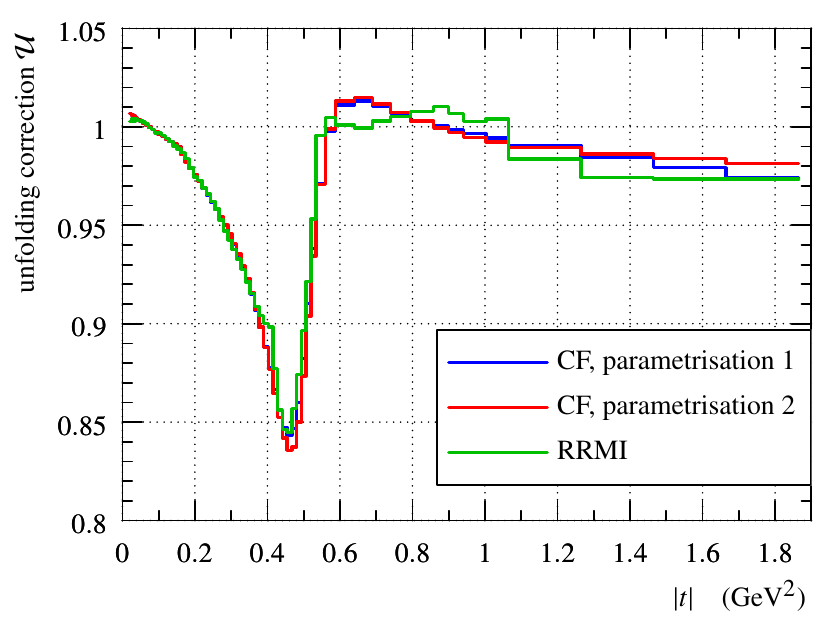}
%\vskip-5mm
\caption{Unfolding correction, ${\cal U}$, as a function of $|t|$. The different colours correspond to various determination techniques, see text.}
\label{f:unfolding}
\end{center}
\end{figure}

Because of the richer structure of the differential cross-section in the full $|t|$ range, the unfolding of resolution effects is more complex than in Ref.~\cite{totem-8tev-90m}. Consequently, an alternative determination is used besides the original method. The original method (denoted ``CF'' in the present article) consists of fitting the observed $t$-distribution with a smooth curve, which serves as an input to a Monte Carlo simulation. This is performed once with and once without simulating the scattering-angle resolution. The ratio of the output histograms gives a set of per-bin corrections factors. Applying them to the yet uncorrected differential cross-section yields a better estimate of the true $t$-distribution and serves as an input to the next iteration. The iterations stop when the difference between the input and output $t$-distributions is negligible (below $0.1\un{\%}$), typically after two iterations. The alternative method performs a regularised resolution-matrix inversion (denoted ``RRMI''), adapted from Chapter~11 in Ref.~\cite{Cowan2002}. The regularisation is needed since the inverted resolution matrix tends to over-amplify statistical fluctuations. It is implemented via minimisation of $\chi^2$ which receives two contributions: one corresponding to the exact resolution-matrix inversion and one proportional to the integral of ${\d^2\over \d |t|^2} \log {\d\sigma\over \d t}$ over the full $|t|$ range. A result comparison is given in Fig.~\ref{f:unfolding}, where the blue and red curves correspond to different parametrisations of the smoothing fit. The red curve is used for correcting the differential cross-section, the others contribute to the systematic-uncertainty estimate.

\begin{figure*}
\begin{center}
\includegraphics{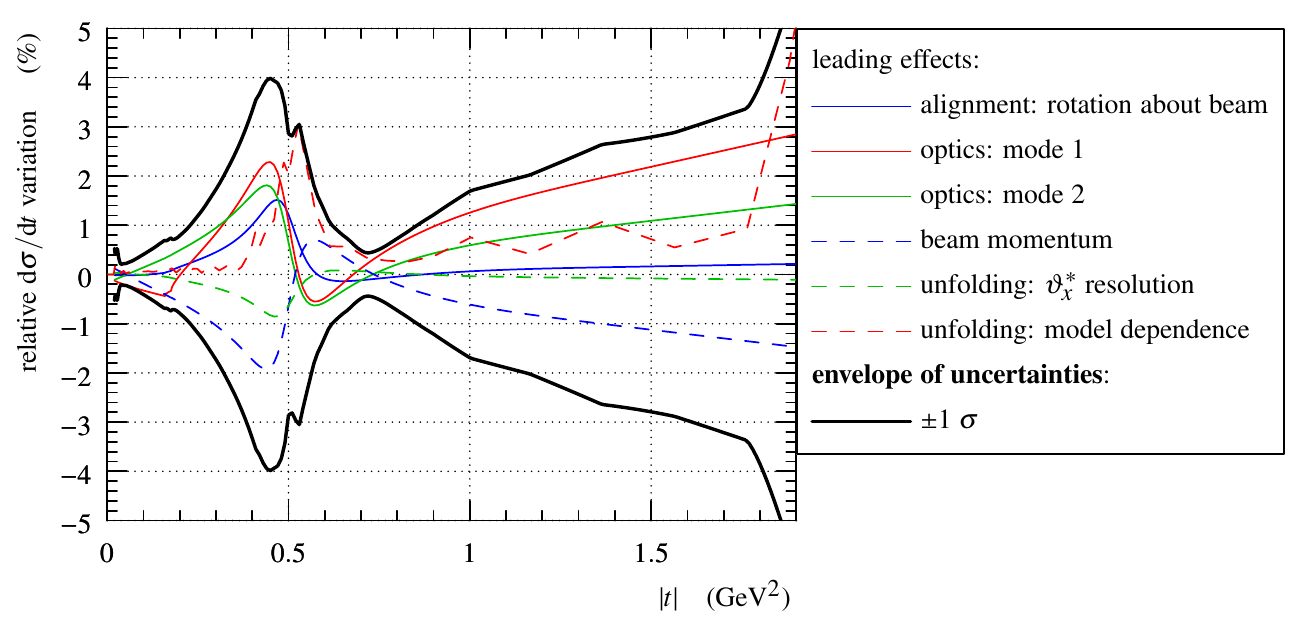}
%\vskip-5mm
\caption{Impact of $t$-dependent systematic effects on the differential cross-section. Each colour curve corresponds to a systematic perturbation at $1\un{\sigma}$. The two contributions due to optics imperfections correspond to the two vectors in Eq.~(8) in Ref.~\cite{totem-8tev-90m}. The thick black envelope is determined by summing all contributions, except normalisation, in quadrature for each $|t|$ value.}
\label{f:syst}
\end{center}
\end{figure*}

The systematic uncertainties considered include:
\begin{itemize}[topsep=0pt,itemsep=-2pt]
\item alignment: RP horizontal and vertical shifts, rotation about beam axis,
\item optics calibration,
\item acceptance correction: uncertainty of the resolution parameters including their possible left-right asymmetry and non-gaussian distribution,
\item uncertainties of the efficiency estimate,
\item uncertainty of the beam momentum~\cite{beam-mom-unc},
\item unfolding: method and fit dependence, uncertainty of the resolution parameters including their full time variation,
\item uncertainty of the normalisation~\cite{totem-8tev-tot1}.
\end{itemize}
The systematic uncertainties were propagated to the differential cross-section using a Monte-Carlo simulation where the correlations between the diagonals were taken into account. The leading systematic effects are evaluated in Fig.~\ref{f:syst}.

\begin{figure}
\begin{center}
\includegraphics{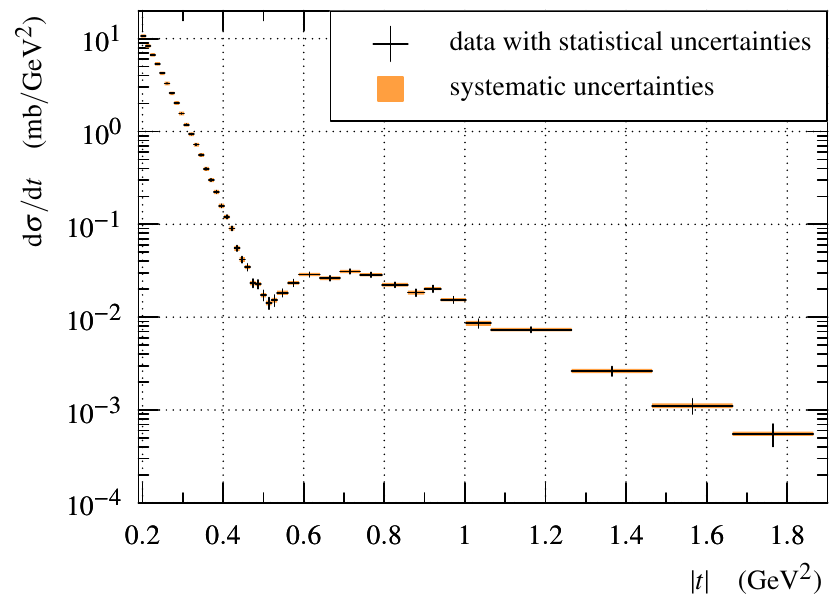}
%\vskip-5mm
\caption{Differential cross-section from Tab.~\ref{t:dsdt}.}
\label{f:dsdt}
\end{center}
\end{figure}

The final differential cross-section with its uncertainties is presented in Tab.~\ref{t:dsdt} and plotted in Fig.~\ref{f:dsdt}.

\begin{table}
\caption{Differential cross-section determined in the present analysis. The left-most two columns describe the $|t|$ bin. The right-most three columns describe the differential cross-section value, statistical and systematic uncertainty. }
\label{t:dsdt}
\begin{center}
%\tiny
\scriptsize
\begin{tabular}{cccccc}
\hline
\multispan2\hss bin $|t|\quad\unt{GeV^2}$\hss &\multispan3 \hss $\d\sigma/\d t\quad \unt{mb/GeV^2}$ \hss\cr
left edge & right edge & value & stat.~unc & syst.~unc\cr
\hline
$0.19595$ & $0.20778$ & $10.697\S\S\S\S$ & $\S0.0433\S\S$ & $\S0.455\S\S\S$\cr
$0.20778$ & $0.21956$ & $\S8.3899\S\S\S$ & $\S0.0395\S\S$ & $\S0.363\S\S\S$\cr
$0.21956$ & $0.23130$ & $\S6.7066\S\S\S$ & $\S0.0363\S\S$ & $\S0.290\S\S\S$\cr
$0.23130$ & $0.24309$ & $\S5.3454\S\S\S$ & $\S0.0330\S\S$ & $\S0.231\S\S\S$\cr
$0.24309$ & $0.25489$ & $\S4.2402\S\S\S$ & $\S0.0300\S\S$ & $\S0.184\S\S\S$\cr
$0.25489$ & $0.26685$ & $\S3.2943\S\S\S$ & $\S0.0267\S\S$ & $\S0.146\S\S\S$\cr
$0.26685$ & $0.27880$ & $\S2.5763\S\S\S$ & $\S0.0240\S\S$ & $\S0.115\S\S\S$\cr
$0.27880$ & $0.29081$ & $\S2.0313\S\S\S$ & $\S0.0216\S\S$ & $\S0.0911\S\S$\cr
$0.29081$ & $0.30293$ & $\S1.5650\S\S\S$ & $\S0.0191\S\S$ & $\S0.0716\S\S$\cr
$0.30293$ & $0.31504$ & $\S1.1847\S\S\S$ & $\S0.0168\S\S$ & $\S0.0560\S\S$\cr
$0.31504$ & $0.32735$ & $\S0.93458\S\S$ & $\S0.0150\S\S$ & $\S0.0436\S\S$\cr
$0.32735$ & $0.33960$ & $\S0.72239\S\S$ & $\S0.0134\S\S$ & $\S0.0338\S\S$\cr
$0.33960$ & $0.35196$ & $\S0.55840\S\S$ & $\S0.0118\S\S$ & $\S0.0261\S\S$\cr
$0.35196$ & $0.36442$ & $\S0.39604\S\S$ & $\S0.00998\S$ & $\S0.0199\S\S$\cr
$0.36442$ & $0.37705$ & $\S0.30024\S\S$ & $\S0.00870\S$ & $\S0.0151\S\S$\cr
$0.37705$ & $0.38962$ & $\S0.22357\S\S$ & $\S0.00756\S$ & $\S0.0114\S\S$\cr
$0.38962$ & $0.40238$ & $\S0.15845\S\S$ & $\S0.00635\S$ & $\S0.00854\S$\cr
$0.40238$ & $0.41514$ & $\S0.12039\S\S$ & $\S0.00555\S$ & $\S0.00636\S$\cr
$0.41514$ & $0.42806$ & $\S0.090421\S$ & $\S0.00480\S$ & $\S0.00462\S$\cr
$0.42806$ & $0.44103$ & $\S0.055239\S$ & $\S0.00375\S$ & $\S0.00337\S$\cr
$0.44103$ & $0.45404$ & $\S0.041774\S$ & $\S0.00326\S$ & $\S0.00243\S$\cr
$0.45404$ & $0.46718$ & $\S0.034636\S$ & $\S0.00297\S$ & $\S0.00177\S$\cr
$0.46718$ & $0.48036$ & $\S0.023328\S$ & $\S0.00245\S$ & $\S0.00134\S$\cr
$0.48036$ & $0.49360$ & $\S0.022714\S$ & $\S0.00246\S$ & $\S0.00107\S$\cr
$0.49360$ & $0.50694$ & $\S0.017260\S$ & $\S0.00218\S$ & $\S0.000881$\cr
$0.50694$ & $0.52039$ & $\S0.014228\S$ & $\S0.00202\S$ & $\S0.000857$\cr
$0.52039$ & $0.53391$ & $\S0.015219\S$ & $\S0.00213\S$ & $\S0.000914$\cr
$0.53391$ & $0.56091$ & $\S0.018318\S$ & $\S0.00172\S$ & $\S0.000940$\cr
$0.56091$ & $0.58791$ & $\S0.023457\S$ & $\S0.00199\S$ & $\S0.00102\S$\cr
$0.58791$ & $0.64006$ & $\S0.028791\S$ & $\S0.00163\S$ & $\S0.00116\S$\cr
$0.64006$ & $0.69012$ & $\S0.026269\S$ & $\S0.00163\S$ & $\S0.00126\S$\cr
$0.69012$ & $0.73990$ & $\S0.031109\S$ & $\S0.00181\S$ & $\S0.00124\S$\cr
$0.73990$ & $0.79439$ & $\S0.028671\S$ & $\S0.00169\S$ & $\S0.00115\S$\cr
$0.79439$ & $0.85850$ & $\S0.022342\S$ & $\S0.00140\S$ & $\S0.00101\S$\cr
$0.85850$ & $0.89954$ & $\S0.018391\S$ & $\S0.00161\S$ & $\S0.000867$\cr
$0.89954$ & $0.94058$ & $\S0.020262\S$ & $\S0.00171\S$ & $\S0.000761$\cr
$0.94058$ & $1.00264$ & $\S0.015403\S$ & $\S0.00123\S$ & $\S0.000640$\cr
$1.00264$ & $1.06469$ & $\S0.0085912$ & $\S0.000932$ & $\S0.000510$\cr
$1.06469$ & $1.26469$ & $\S0.0073395$ & $\S0.000495$ & $\S0.000306$\cr
$1.26469$ & $1.46469$ & $\S0.0026302$ & $\S0.000309$ & $\S0.000141$\cr
$1.46469$ & $1.66469$ & $\S0.0011130$ & $\S0.000208$ & $\S0.000063$\cr
$1.66469$ & $1.86469$ & $\S0.0005551$ & $\S0.000151$ & $\S0.000029$\cr
\hline
\end{tabular}
\end{center}
\end{table}

%----------------------------------------------------------------------------------------------------
\section{Characterisation of the dip-bump structure}
\label{s:dip-bump}

Two complementary kinds of fits are used to extract parameters of the dip and the bump. In both cases, the fits are performed by minimising the standard binned $\chi^2$ where only the statistical uncertainties are considered. Evaluation of various systematic uncertainties is described later in this section.

``{\bf Local fits}'' represent parabolic fits performed through local neighbourhoods of the dip and the bump. This choice corresponds to the lowest Taylor approximation of the differential cross-section shape locally in the areas of interest which leads to a satisfactory description of the data. It can thus be regarded as the most model-independent approach. The dip fit is carried over the range $0.47 < |t| < 0.56\un{GeV^2}$ and gives $\chi^2/\hbox{ndf} = 0.60$. The bump fit goes through $0.56 < |t| < 0.86\un{GeV^2}$ and yields $\chi^2/\hbox{ndf} = 1.99$. 

``{\bf Global fit}'' represents a single fit throughout the dip-bump region according to the parametrisation from Eq.~(1) in Ref.~\cite{d0-totem-prl}:
\begin{equation}
\label{eq:global-param}
{\d\sigma\over\d t} = p_0\,\exp\left( p_1\, |t| + p_2\,|t|^2 \right) + p_3\,\exp\left( p_4\, |t| + p_5\,|t|^2 + p_6\,|t|^3 \right) \ .
\end{equation}
This parametrisation has been shown to describe well the trend present also at other LHC energies ($\sqrt s = 2.76$, $7$ and $13\un{TeV}$). It can therefore be considered as carrying some information from the other TOTEM measurements.
% The parametrisation has been shown flexible enough to describe the TOTEM data at all other LHC energies and at the same time rigid enough not to pick up exaggerated fluctuations.
The fit is made over the range $0.42 < |t| < 1.06\un{GeV^2}$ and gives $\chi^2/\hbox{ndf} = 2.22$.

\begin{figure}
%\hbox{}\vskip-7mm
\begin{center}
\includegraphics{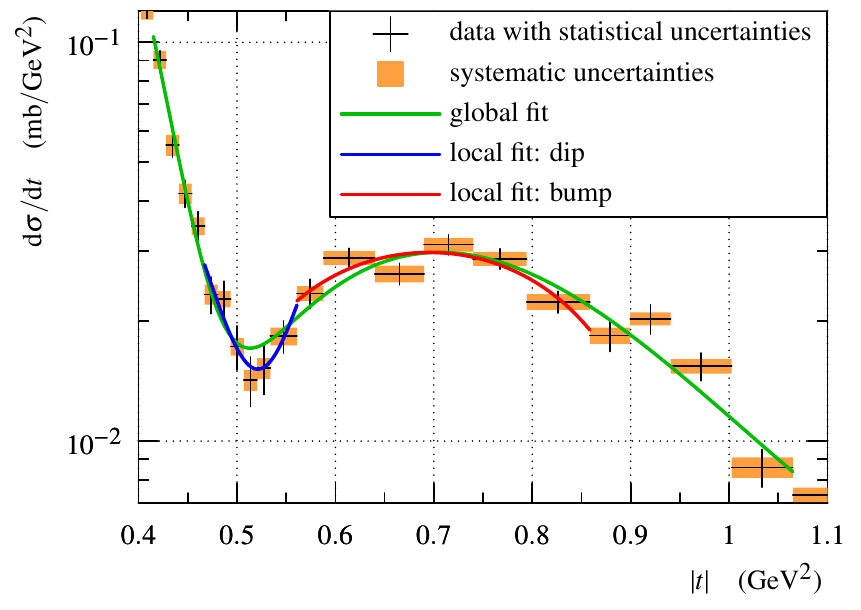}
%\vskip-5mm
\caption{Zoom of the differential cross-section (black points) in the dip-bump region. The coloured curves correspond to the fits discussed in the text.}
\label{f:dip bump fits}
\end{center}
\end{figure}

The fit results are summarised in Table \ref{t:dip_bump} and visualised in Fig.~\ref{f:dip bump fits}. The local fits are considered as the main result, the global fit as a cross-check. The results from the local and global fits are compatible within the uncertainties.

One source of the systematic uncertainties of the dip and bump parameters are the systematic uncertainties of the differential cross-section, cf.~the list at the end of Section \ref{s:dsdt}. For each perturbation from the list, $\d\sigma/\d t$ distributions biased by $\pm1\un{\sigma}$ have been re-fitted and parameter offsets from the unbiased fit have been calculated. The offsets are summed in quadrature from all systematic perturbations for the final estimate shown in the column ``$\d\sigma/\d t$ systematic uncertainty'' in Table~\ref{t:dip_bump}.

Another source of systematic uncertainties is the subjective choice of the fitting range. To evaluate this contribution we have repeated the fits altering the fit range(s) by $\pm 1$ bin at each side of the range (all possible combinations considered). The standard deviation of the fit results has then been considered as the ``range'' systematic uncertainty, cf.~``range uncertainty'' column in Table~\ref{t:dip_bump}.

Yet another source of systematic uncertainties can be related to the subjective choice of the fitting function. This is particularly pertinent to $\d\sigma/\d t|_{\rm dip}$, where the two fits give considerably different results, and $|t|_{\rm bump}$ where the coarse granularity of the bins and their large fluctuations allow for many other fit parametrisations, likely yielding notably different positions of the bump. These contributions are summarised in the ``parametrisation uncertainty'' column in Table~\ref{t:dip_bump}.

Regarding $\d\sigma/\d t|_{\rm dip}$, the difference in the two fit results can be interpreted as follows. The local fit extracts a cross-section value close to the central bin values present in the data. In contrary, the ``rigidity'' of the global fit does not allow for such a deep dip -- indeed, at other LHC energies the dip seems less pronounced than at $8\un{TeV}$ \cite{d0-totem-prl}. Therefore, in order to cover for the possibility that the truth is better expressed by the global fit, we assign an additional uncertainty to $\d\sigma/\d t|_{\rm dip}$ of $2.0\un{\mu b/GeV^2}$, which corresponds to the difference between the two fits.

Regarding $|t|_{\rm bump}$, we attribute an additional uncertainty of $0.005\un{GeV^2}$ so that the full uncertainty becomes about the half size of the bin -- which seems to be a natural limit for the precision in $|t|$ of the bump, given the large statistical fluctuations in $\d\sigma/\d t$ of the bins in the region.

The depth of the dip can be evaluated with a ratio:
\begin{equation}
R = {\d\sigma/\d t|_{\rm bump}\over \d\sigma/\d t|_{\rm dip}}\ .
\end{equation}
The numerator-denominator correlations are taken into account in the uncertainty estimates presented in Table~\ref{t:dip_bump}.

\begin{table*}
\caption{Dip and bump parameters as extracted by the local fits (central part) and the global fit (right-most column) techniques. The uncertainties quoted for the local fits are (from left to right): statistical, systematic (propagated from $\d\sigma/\d t$ analysis), due to the variation of the fit range, due to the fit parametrisation choice and the full uncertainty (quadratic combination of the preceding four contributions). There are no uncertainties quoted for the global fit since it is only used to cross-check the results with an alternative parametrisation.}
\label{t:dip_bump}
\begin{center}
%\scriptsize
\begin{tabular}{cc|cccccc|c}
\hline
                                              &                        &\multispan6 \hfil local fits\hfil\vrule                                                  & global fit\cr
quantity                                      & unit                   & central      & statistical  & $\d\sigma/\d t$ systematic  & range        & parametrisation & full         & central\cr
                                              &                        & value        & uncertainty  & uncertainty  & uncertainty  & uncertainty  & uncertainty  & value\cr
\hline
$|t|_{\rm dip}$                               & $\rm GeV^2$            &        0.521 &        0.005 &        0.001 &        0.004 &        0.000 &        0.007 &        0.513\cr
$\left.{\d\sigma\over\d t}\right|_{\rm dip}$  & $\rm\mu b\over GeV^2$  &         15.1 &          1.2 &          0.8 &          0.4 &          2.0 &          2.5 &         17.1\cr\hline
$|t|_{\rm bump}$                              & $\rm GeV^2$            &        0.695 &        0.010 &        0.003 &        0.023 &        0.005 &        0.026 &        0.706\cr
$\left.{\d\sigma\over\d t}\right|_{\rm bump}$ & $\rm\mu b\over GeV^2$  &         29.7 &          1.1 &          1.3 &          0.7 &          0.0 &          1.8 &         29.7\cr\hline
$R$                                           &                        &         1.96 &         0.18 &         0.05 &         0.07 &         0.26 &         0.33 &         1.74\cr
\hline
\end{tabular}
\end{center}
\end{table*}

%----------------------------------------------------------------------------------------------------
\section{Summary}
\label{s:summary}

The TOTEM collaboration has measured the differential cross-section of elastic pp scattering at $\sqrt{s} = 8\,$TeV in the range $0.2\un{GeV^{2}} < |t| < 1.9\un{GeV^{2}}$, complementing the previously published result \cite{totem-8tev-90m} for $0.03\un{GeV^{2}} < |t| < 0.2\un{GeV^{2}}$ on the basis of the same data set.
The new measurement confirms the presence of the dip-bump structure also observed at the energies $\sqrt{s}$ = 2.76, 7 and 13\,TeV \cite{totem-2.76tev-diff,totem-7tev-3.5m,totem-13tev-diff}. The detailed qualification of this structure allows an extrapolation of its characteristics to the Tevatron energy of 1.96\,TeV and thus a quantitative comparison with the p$\bar{\rm p}$ measurement by the D0 experiment \cite{d0-totem-prl}.

%----------------------------------------------------------------------------------------------------
\section*{Acknowledgments}

The TOTEM collaboration is grateful to the CERN beam optics development team for the design and the successful commissioning of the special $\beta^* = 90\un{m}$ optics and to the LHC machine coordinators for scheduling the dedicated fill. We acknowledge the support from the collaborating institutions and also NSF (US), the Magnus Ehrnrooth Foundation (Finland), the Waldemar von Frenckell Foundation (Finland), the Academy of Finland, the Finnish Academy of Science and Letters (The Vilho, Yrj\"o and Kalle V\"ais\"al\"a Fund), the Circles of Knowledge Club (Hungary), the NKFIH/OTKA grant K 133046 and the Human Resources Development Operational Programme (EFOP)  grant No. 3.61-16-2016-00001 (Hungary). Individuals have received support from Nylands nation vid Helsingfors universitet (Finland), M\v SMT \v CR (the Czech Republic), the J\' anos Bolyai Research Scholarship of the Hungarian Academy of Sciences, the New National Excellence Program of the Hungarian Ministry of Human Capacities and the Polish Ministry of Science and Higher Education Grant No. MNiSW DIR/WK/2018/13.

%----------------------------------------------------------------------------------------------------
\bibliographystyle{h-elsevier.bst}
\bibliography{bibliography}

\end{document}